\documentclass[aps,preprint,prb,amsmath,amssymb,showpacs,superscriptaddress]{revtex4-1}

\usepackage{graphicx}
\usepackage{color}

\makeatletter

\@ifundefined{textcolor}{}
{
 \definecolor{BLACK}{gray}{0}
 \definecolor{WHITE}{gray}{1}
 \definecolor{RED}{rgb}{1,0,0}
 \definecolor{GREEN}{rgb}{0,1,0}
 \definecolor{BLUE}{rgb}{0,0,1}
 \definecolor{CYAN}{cmyk}{1,0,0,0}
 \definecolor{MAGENTA}{cmyk}{0,1,0,0}
 \definecolor{YELLOW}{cmyk}{0,0,1,0}
}

\newcommand{\etal}{{\it et al.}}

\begin{document}

\title{Strong interaction between electrons and collective excitations in multiband superconductor MgB$_2$}

\author{Daixiang Mou}
\author{Rui Jiang}
\author{Valentin Taufour}
\author{Rebecca Flint}
\author{S. L. Bud'ko}
\author{P. C. Canfield}
\affiliation{Division of Materials Science and Engineering, Ames Laboratory, U.S. DOE, Ames, Iowa 50011, USA}
\affiliation{Department of Physics and Astronomy, Iowa State University, Ames, Iowa 50011, USA}
\author{J.~S.~Wen}
\affiliation{Condensed Matter Physics and Materials Science Department, Brookhaven
National Laboratory, Upton, New York 11973, USA}
\author{Z.~J.~Xu}
\affiliation{Condensed Matter Physics and Materials Science Department, Brookhaven
National Laboratory, Upton, New York 11973, USA}
\author{Genda~Gu}
\affiliation{Condensed Matter Physics and Materials Science Department, Brookhaven
National Laboratory, Upton, New York 11973, USA}
\author{Adam Kaminski}
\affiliation{Division of Materials Science and Engineering, Ames Laboratory, U.S. DOE, Ames, Iowa 50011, USA}
\affiliation{Department of Physics and Astronomy, Iowa State University, Ames, Iowa 50011, USA}
date{Today}

\date{\today}    

\begin{abstract}
We use a tunable laser ARPES to study the electronic properties of the prototypical multiband BCS superconductor MgB$_2$. Our data reveal a strong renormalization of the dispersion (kink) at $\sim$65 meV, which is caused by coupling of electrons to the E$_{2g}$ phonon mode. In contrast to cuprates, the 65 meV kink in MgB$_2$ does not change significantly across T$_c$. More interestingly, we observe strong coupling to a second, lower energy collective mode at binding energy of 10 meV. This excitation vanishes above T$_c$ and is likely a signature of the elusive Leggett mode. 
\end{abstract}
\maketitle

In conventional superconductors, the pairing is mediated by phonons and favored by strong electron-phonon coupling, as described by Bardeen-Cooper-Schrieffer (BCS) theory\cite{Bardeen}. This strong electron-phonon coupling in general gives rise to a renormalization of the band dispersion called a ``kink'' and an abrupt change of quasiparticle lifetime at an energy related to the phonon frequency, $\Omega$.  This idea has been extended to unconventional superconductors, where the mechanism of pairing is unknown, and the coupling of electrons to several collective excitations was reported \cite{MODE1,MODE2,Valla,Bogdanov,Kaminski,KondoLowKink}. Their origin and relation to pairing is still debated. 
For example the "70meV" kink in Bi2212 cuprate is strongest at the antinode and vanishes above T$_c$\cite{Kaminski} a behavior that resembles the magnetic resonance mode reported by inelastic neutron scattering \cite{CuRes}; however, the electron-phonon interaction may have similar characteristics \cite{TCuk, Devereaux}. On the other hand, the kinks in the dispersion along the nodal direction in single layer Bi2201 do not seem to change significantly with temperature \cite{Meevasana}.
Surprisingly, there is little data on dispersion renormalization effects in conventional superconductors; several low energy kinks have been reported in NbSe$_2$ \cite{Rossnagel}, but the situation is complicated by the presence of a charge density wave phase coexisting with superconductivity, and the measurements were carried out only at low temperatures. In other materials, difficulties in clearly observing three dimensional band dispersions and low transition temperatures are limiting factors. MgB$_2$ \cite{Nagamatsu,Budko,Kortus,JMAn,AYLiu,Choi} is a notable exception:
it is a layered material with multi-gap, phonon-mediated superconductivity at T$_c$= 39K with some quasi-two-dimensional bands, making it an ideal candidate to study the temperature dependence of the dispersion renormalization due to electron-phonon coupling.  LDA calculations \cite{Kortus} predict four bands crossing Fermi level in MgB$_2$: two quasi 2D $\sigma$-bands from $p_x$, $p_y$ orbitals around $\Gamma$ and two 3D $\pi$-bands from $p_z$ orbital \cite{Uchiyama,Souma, Tsuda}. The superconductivity is believed to caused by the $E_{2g}$ phonon mode at 75meV that couples strongly to the 2D $\sigma$-bands, but more weakly to the $\pi$-bands \cite{AYLiu}, leading to two different gaps, $\Delta_\sigma = 6.5$ meV and $\Delta_\pi = 1.5$meV, as revealed by previous tunneling \cite{Szabo} and ARPES studies \cite{Souma}.  Inelastic neutron scattering studies have reported optical phonon modes at $\sim$35, 55, 75, 85 and 100 meV \cite{OsbornPRL2001}, while Raman scattering reports a single, broad asymmetric peak at 75meV, attributed to the $E_{2g}$ mode, as well as a sharp peak at $2\Delta_\sigma = 12$meV due to a pair breaking excitation \cite{Quilty}. As a bonus, due to its multi-gap nature, MgB$_2$ also contains another exotic collective mode that can couple to the electrons: the Leggett mode \cite{Leggett,Agterberg, Brinkman, Ponomarev, Blumberg}. This mode is a longitudinal fluctuation corresponding to equal and opposite displacements of the two condensates.  As it is a neutral excitation, it is not pushed up to the plasma frequency like the Bogoliubov-Anderson mode, however, it is predicted to have a mass in-between $2\Delta_\pi$ and $2\Delta_\sigma$ for MgB$_2$, where it will be partially damped by decaying into $\pi$ quasiparticles \cite{LMtheory}. As the superconductivity is believed to be phonon mediated and due to intraband pairing, the two gaps are expected to have the same relative sign, giving rise to $s_{++}$ pairing, by contrast to the iron based superconductors, where interband repulsive pairing likely leads to $s_{\pm}$ pairing and a Leggett mode is not expected.

High resolution band dispersion data demonstrating the coupling of the conduction electrons to collective excitations such as phonons or Leggett modes are not available in the published literature.  One of the main reasons is that high quality single crystals need to be synthesized under high pressure, resulting in rather small $\le$500 $\mu$m crystals, which are difficult to measure in traditional ARPES setups. We use tunable laser ARPES to study the electronic properties of MgB$_2$ multiband superconductivity. The use of low photon energy increases significantly the bulk sensitivity due to increased escape depth and momentum resolution due to increased {\AA}$^{-1}/$deg ratio. The ability to focus the laser beam down to $\sim$30$\mu$m enables measurement of very small single crystals and also helps to improve the momentum resolution. We find evidence for strong coupling of conduction electrons to a 75 meV acoustic phonon with $\lambda$ estimated at $\sim1.3$ that persists above T$_c$, unlike those in the cuprates. In fact we observe no significant changes with temperature up to 65K, more than 50\% above T$_c$. Furthermore, we observe a very strong renormalization of the dispersion in the superconducting state at $\sim$ 10 meV. Instead of the expected Bogoliubov-like back bending of the dispersion, a sharp, non-dispersive quasiparticle peak centered at the gap energy of 6.5 meV is present and is separated by a dip from the high energy spectral weight. All these features vanish above T$_c$ as expected for a superfluid excitation and are likely due to interaction of electrons with a Leggett mode.

MgB$_2$ single crystals with T$_c$ = 39 K and typical size of $0.5\times0.5\times0.3$ mm$^3$ were grown in Ames Laboratory by a high pressure synthesis technique similar to that outlined in Ref. \cite{Karpinski} using pure $^{11}$B isotope. Optimally doped Bi$_2$Sr$_2$CaCu$_2$O$_{8+\delta}$ (Bi2212) single crystals with $T_{\rm c}$=93K (OP93K) were grown by the conventional floating-zone (FZ) technique. Sample were cleaved \emph{in situ} at a base pressure lower than 8 $\times$ 10$^{-11}$ Torr. ARPES measurements were carried out using a laboratory-based system consisting of a Scienta R8000 electron analyzer and tunable VUV laser light source\cite{RJiang}. All data were acquired using a photon energy of 6.7 eV. The energy resolution of the analyzer was set at 1 meV and angular resolution was 0.13$^\circ$ and $\sim$ 0.5$^\circ$ along and perpendicular to the direction of the analyzer slits, respectively. Samples were cooled using a closed cycle He refrigerator and temperature was measured using a silicon-diode sensor mounted on the sample holder. The energy corresponding to the chemical potential was determined from the Fermi edge of a polycrystalline Au reference in electrical contact with the sample. The aging effect was checked by recycle measurements. The consistency of the data was confirmed by measuring several samples.

\begin{figure}[htbp]
\centering
\includegraphics[width=6in]{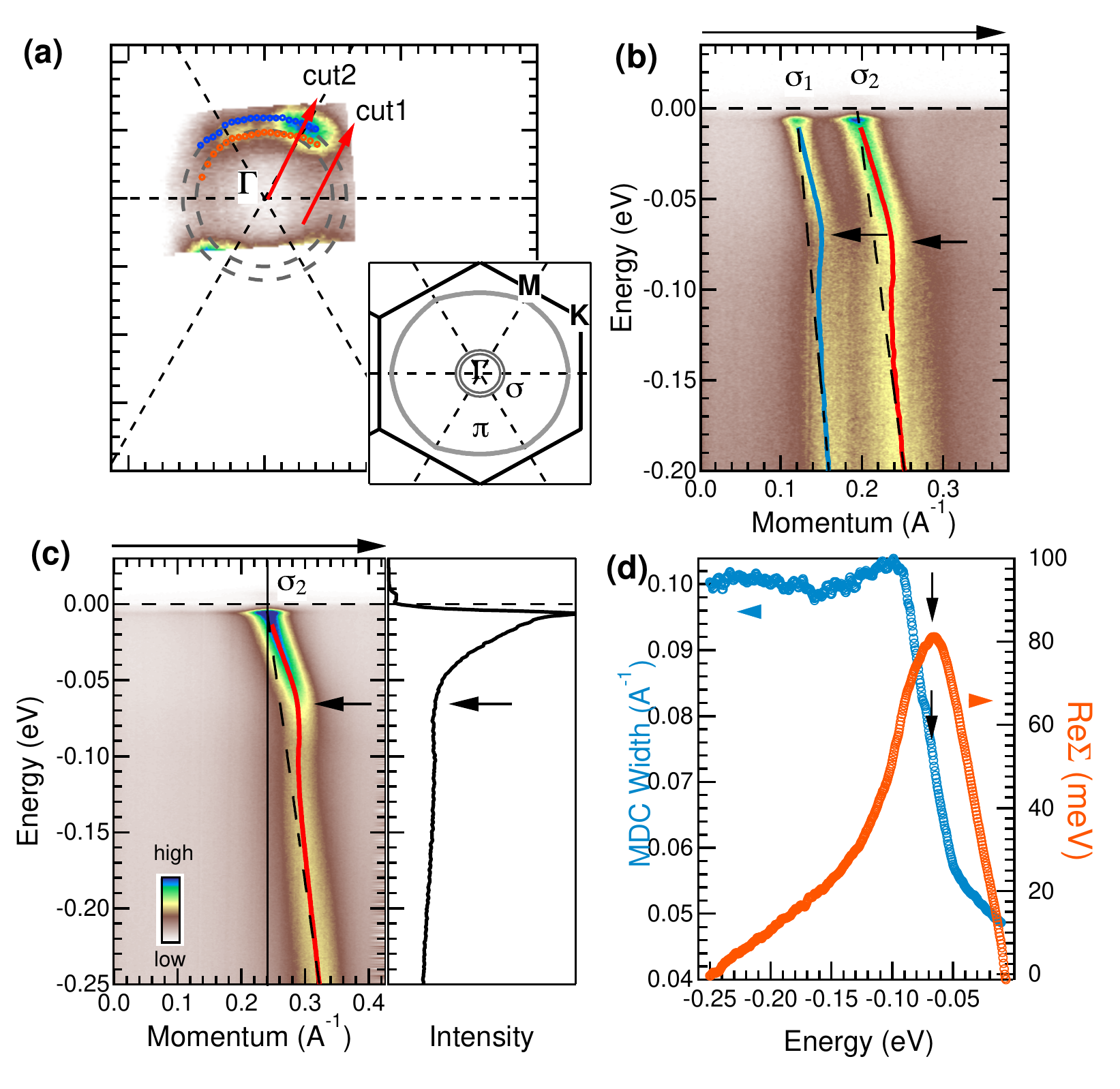}
\caption{(Color online) Renormalization effects (kink) in $\sigma$-band. (a) Measured FS at 40K. Blue and red circles mark the Fermi momentum extracted from MDC peak position at E$_F$. Two red arrows mark locations of measured cuts in the momentum space. Insert is a schematic diagram of the FS and Brillouin zone.  (b)  ARPES intensity along cut \#1 at 15K. Solid lines are dispersions obtained by MDC fits for the two bands. Black dashed lines signify bare band dispersion extrapolated from higher binding energies. (c) left: same as (b) but along cut \#2; right:  EDC at K$_F$ for outer $\sigma$-band. (d) Energy dependence of the MDC width (blue solid line) and effective real part of self-energy (red dotted line) for data in panel (c). Arrows mark the energy location of the kink and associated features.}
\end{figure}


The Fermi surface data are shown in Fig. 1a, with a schematic plot of the Brillouin zone inset. We used the peak position of MDCs at E$_F$ to quantitatively extract k$_F$. Results are superposed as red points on image data. Both $\sigma$ FS sheets are round with $\mid$k$_F$$\mid$  $\sim$ 0.2 {\AA}$^{-1}$ and $\sim$ 0.25 {\AA}$^{-1}$ respectively. If we ignore the small warping of these two $\sigma$ sheets along k$_z$, the contribution to carrier concentration would be 0.069 holes for inner FS sheet and 0.108 holes for outer one. The measured area of the outer FS is consistent with previous quantum oscillation results, while the inner one is slightly larger\cite{Yelland}. 
Fig. 1 (b, c) show ARPES intensity plots at two different cuts in the Brillouin zone. In the data measured along cut \#1 (Fig. 1b) the Fermi crossing for both $\sigma$ sheets are clearly visible, whereas in the cut \#2, along symmetry axis (Fig. 1c), intensity of the inner $\sigma$-band is strongly suppressed due to the matrix element effect.  In order to quantitatively analyze the renormalization effects due to the collective modes, we fit the MDCs of each data set with Lorenzians and plot such extracted dispersion as blue and red lines in Fig. 1 (b) and (c). It should be noted that MDC peaks do not reflect the dispersion at very low energies in the presence of the superconducting gap therefore the fitting is carried out only for binding energies larger than $\sim$2$\Delta$. We estimate bare dispersion by extrapolation from higher binding energies and plot them as dashed lines. In all three  data sets, a very pronounced kink structure is clearly visible (indicated by arrows), where the renormalized dispersion deviates from bare estimate. In this case, the renormalization of the dispersion (kink) is the fingerprint of coupling the conduction electrons to phonon mode(s). This coupling is rather strong and the energy distribution curves (EDCs) at k$_F$ develop a dip (right side of panel 1 (c)). The kink in dispersion and dip in EDC occurs at $\sim$ -(70-75) meV. By comparison with Raman data \cite{Quilty}, we can conclude that it is due to the E$_{2g}$ optical phonon responsible for pairing in MgB$_2$ \cite{AYLiu}.  We subtract an estimated bare dispersion from the measured one to obtain the approximate $\Sigma'(\omega)$, and use the widths of the Lorentzian MDC fits to obtain $\Sigma''(\omega)$. These quantities are plotted in panel 1 (d).  Not surprisingly, $\Sigma'(\omega)$ has a peak at -70 meV, which is very close to the phonon energy. $\Sigma''(\omega)$ rises rapidly with binding energy and has a mid-point of the step roughly located at the same energy as the peak in $\Sigma'(\omega)$. The line shape of both curves is consistent with them being related by Kramers-Kronig. This is further verified by simulations presented in Supplemental Material.

\begin{figure}[htbp]
\centering
\includegraphics[width=5in]{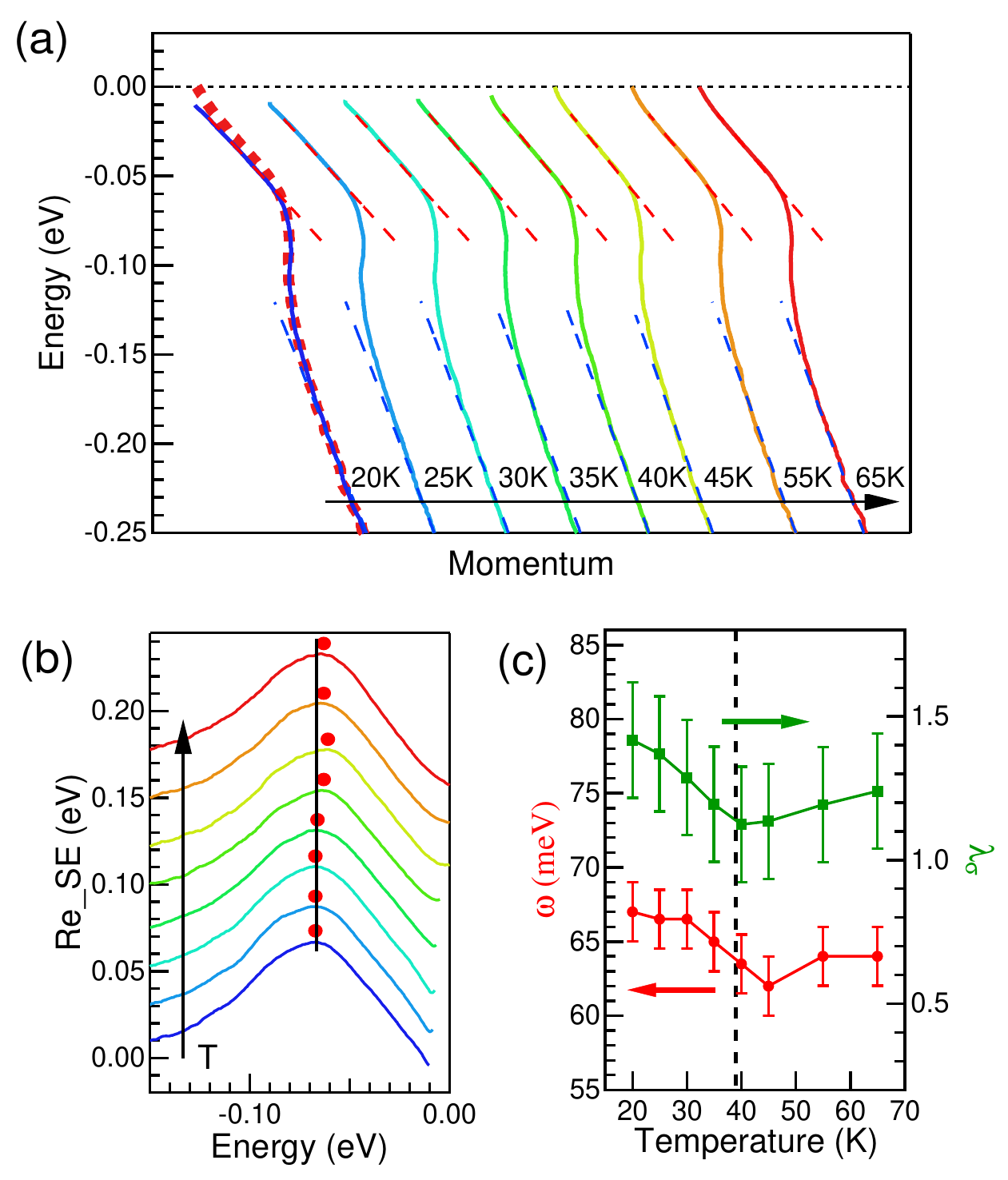}
\caption{(Color online) Temperature dependence of the renormalization effect. (a) Band dispersions obtained from MDC fits along cut \#2 (see Fig. 1a) measured at different temperatures. Data are shifted horizontally for clarity. Band dispersion at 25K (blue solid line) is superimposed with  65K data (thick red doted line) for direct comparison of the effect below and above T$_c$. Blue doted and red dashed lines illustrate the dispersion used to extract velocity at low and high binding energies for estimating coupling constant. (b) Extracted effective real part of self-energy obtained from (a). Data are shifted vertically with 25meV interval for each temperature. Red dots mark peak position. (c) Temperature dependence of kink energy (red squares) and coupling constant (green circles).}
\end{figure}

The dispersion data extracted by fitting the positions of the MDC peaks for several temperatures both below and above T$_c$ are shown in Fig. 2(a).  There are no significant changes to the kink structure across T$_c$ as evident from overlay of the high temperature curve (dashed red line) onto the lowest temperature one. The only noticeable changes occur at low energies  due to the opening of the superconducting gap.   Following the procedure outlined above, we extract $\Sigma'(\omega)$ for each temperature, and plot these in Fig. 2(b).  The peak position does move to lower energies as the temperature decreases, starting at T$_c$, consistent with the expectation that the peak frequency increases from $\Omega \rightarrow \Omega +\Delta$.  However, the magnitude of the shift, $3$meV is significantly smaller than the expected $\Delta_\sigma = 6.5$meV.  In fact, this shift should  be even larger as  the screening electrons are gapped out and E$_{2g}$ phonon hardens below T$_c$. This hardening was predicted to be $\approx 10\%$ or  $\approx 7$meV~\cite{AYLiu}, but Raman measurements find that it only shifts by $\Delta \Omega \approx 2.5$meV~\cite{Mialitsin}.  So the overall shift below T$_c$ is naively expected to be $\Delta_\sigma + \Delta \Omega \approx 9$meV.  However, this analysis neglects the multi-gap nature, which may account for shifts as small as $3$meV by allowing scattering into the $\pi$ band.

In addition to extracting the energy of the collective mode, we also can estimate the electron-phonon coupling $\lambda =\nu_0/\nu_F-1$, where $\nu_0$ and $\nu_F$ are the bare and renormalized Fermi velocity respectively.  $\nu_0$ is estimated from the dispersion at high energy and $\nu_F$ is obtained from the dispersion above the kink (as illustrated in Fig. 2a by red dashed and blue dotted lines). For cut \#2 in Fig. 1c, $\nu_0\sim$3.77$eV{\AA}$ and $\nu_F\sim$1.56$eV{\AA}$, which implies $\lambda_\sigma$=1.42.  The electron-phonon coupling can be calculated numerically, where it results from scattering between the  $\sigma$ and $\pi$-bands, $\lambda_{\sigma \pi} = .23$ or from intra-$\sigma$-band scattering, $\lambda_{\sigma\sigma} = .96$, for a total of $\lambda_\sigma = 1.19$\cite{Mazin,AYLiu}, which is slightly smaller, but within error bars of our results given uncertainty of estimating $\nu_0$.  

\begin{figure}[htbp]
\centering
\includegraphics[width=6in]{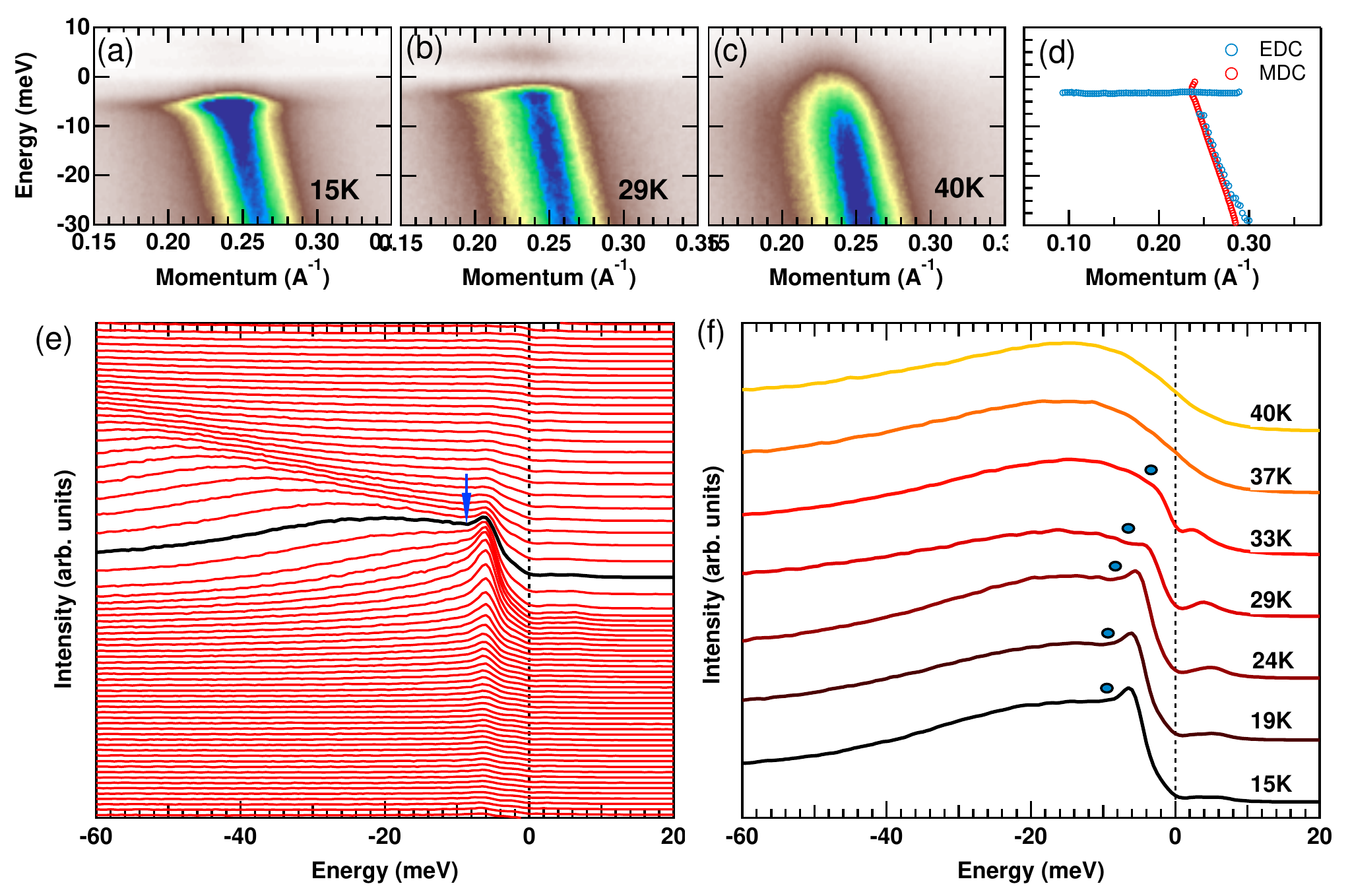}
\caption{(Color online) Low energy band dispersion and kink. (a-c) ARPES intensity measured at 15K, 29K and 40K along cut \#2 (see Fig. 1a). (d) dispersion at low temperature extracted using MDC and EDC fits to data in (a). (e) EDCs measured at 15K, arrow marks location of the dip which roughly corresponds to the energy of the Leggett mode. (f) temperature dependence of the EDC's. Clear dip in the spectrum due to interaction with Leggett more is marked by an oval.}
\end{figure}

We now turn to describe the second, low energy excitation present only below the superconducting transition. In Fig. 3 (a-c) we plot the dispersion of the $\sigma_1$-band in close proximity to E$_F$ below and above T$_c$. At low temperature, well below T$_c$, the shape of the dispersion is rather unusual for a superconductor. Instead of back bending of the band as it reaches the energy of the superconducting gap, a sharp streak of intensity is present at the gap energy on either side of k$_F$. At 29K (panel b) this feature is also above E$_F$ due to increased thermal excitation. All of these features vanish just above T$_c$ (panel c), where the ordinary conduction band is present. In panel (d) we plot the dispersion of the low temperature features extracted using EDC and MDC fits.  The sharp peak of intensity is almost dispersionless and persists over $\delta$k of 0.2  \AA$^{-1}$ unlike what is expected for Bogolubov quasiparticles \cite{Campuzano, Matsui}. The EDCs for low temperature data are plotted in panel (e). Here again we observe a sharp, dispersion less quasiparticle peak centered at the energy of the SC gap and separated from the rest of the spectral weight by a dip at $\sim$10 meV (marked by an arrow). The temperature dependence of the EDCs slightly off k$_F$ are shown in panel (f). A clear dip in the spectrum is observed, which vanishes as the temperature approaches T$_c$. All of these features are due to abrupt changes in the self energy (i. e. onset of resonant scattering) and are very characteristic of an interaction between the electrons and a collective excitation.
 It should be noted that opening of the SC gap can lead to suppression of scattering within energy of 3$\Delta_\sigma$=19.5 meV for intra-band and $\Delta_\sigma$+2$\Delta_\pi$= 9.5 meV for inter-band electron channels. We do not observe signatures of suppression of intra-band scattering (i. e.  line shape features at 19.5 meV) which are deemed to be stronger than inter-band ones\cite{Mazin2002}. Further more, strong suppression of scattering and reduction of $\Sigma''(\omega)$ below certain energy alone cannot produce a spectral dip and non-dispersive peaks. That requires a resonant process such as interaction with a collective mode that causes a peak in $\Sigma''(\omega)$. An illustration of this in form of simulations of the spectral function for various scenarios of $\Sigma''(\omega)$ are presented in the Supplementary Material.
The lowest energy of an optical phonon in MgB$_2$ is 35 meV and there are no other obvious low energy excitations that could couple strongly to the electrons other than the Leggett mode. 
Whereas Raman spectroscopy has found the Leggett mode at 9.4meV \cite{Blumberg}, the similarity of this value to the dip energy of 10 meV is coincidental, as the dip occurs at the frequency where electrons above the gap may scatter into the collective mode plus quasiparticles above the superconducting gap. Therefore we extract the value of the mode energy from ARPES data of 3.5 meV, which is the difference between energy location of sharp peak and dip in the spectrum. We stress that this is an estimate as the dip location is slightly affected by functional form of the self energy. 
The vertex corrections for Raman spectroscopy\cite{Blumberg,LMtheory} and ARPES are almost certainly different and can potentially explain the the difference between the Leggett mode energy measured by the two techniques. Previous electron spectroscopy studies  reported the value of the Leggett mode energy of 3.9-4.0 meV, consistent with our results. While the bare mode was calculated to have a frequency of 5.1-6.2meV\cite{Blumberg,LMtheory}, if the experimentally measured $\Delta_{\pi} = 1.5$meV is used, the calculated frequencies decrease to 3.8-4.5meV.  More theoretical efforts will be required to fully understand the origin of the difference. 

\begin{figure}[htbp]
\centering
\includegraphics[width=5in]{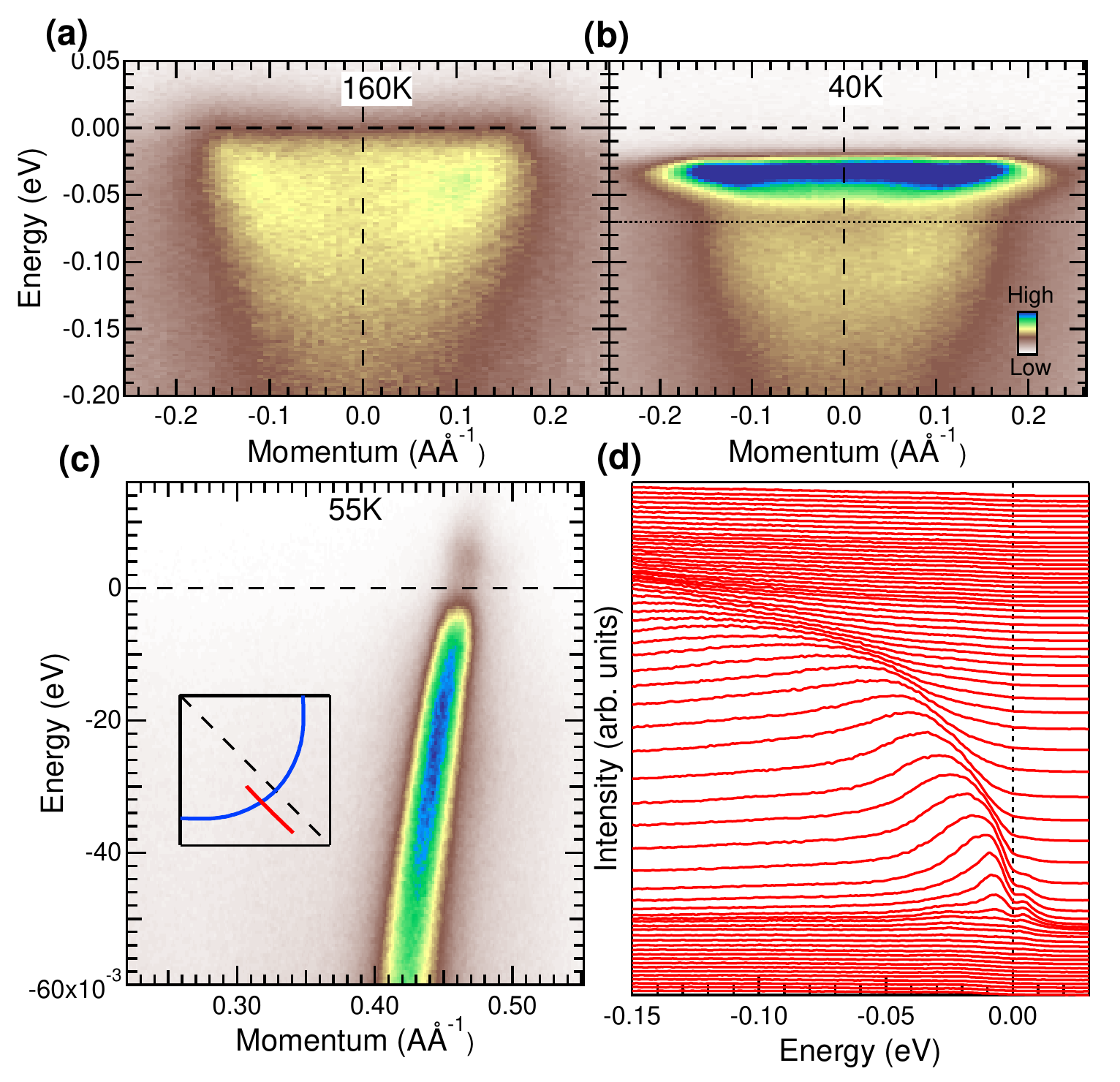}
\caption{(Color online) ARPES data from OP Bi2212 for comparison. Intensity plot at antinode in a)  normal and b) superconducting state measured using Helium discharge lamp and 21.2 eV photons. Dashed line in b) marks location of dip (local minimum of intensity) due to interaction with 40 meV collective mode. This feature is absent in normal state (panel a). c) Intensity plot and d) EDCs  in SC state along cut slightly off the node (marked in inset), where the SC gap magnitude is $\sim$8 meV measured using laser source and 6.7 eV photons. The intensity peak above E$_F$ is due to  electrons thermally excited above 2$\Delta$.  }
\end{figure}

 We also note that  spectral characteristics reported in Fig. 3 (non-dispersive sharp peak and presence of a dip) closely resemble features observed at antinode in the cuprates that are attributed to strong coupling of electrons to a collective mode.  The single order parameter of the cuprates does not support a Leggett mode, but these properties are characteristic of any collective mode developing only below the superconducing transition. In Fig. 4 we show data from optimally doped Bi2212 for comparison. Panels a and b show data at the antinode where electrons couple strongly to a collective mode below T$_c$. In SC state a very sharp, weakly dispersing peak emerges at the energy of the SC gap followed by a dip of intensity. These two features - lack of substantial dispersion of the sharp quasiparticle peak and presence of a dip are key signatures of the interaction with a collective mode. The data close to the node ($\Delta$=8 meV) where the coupling to low energy modes in Bi2212 is very weak\cite{KondoLowKink} in SC state display Bogoliubov-like dispersion and sharp, non-dispersive peaks are absent. This is in stark contrast to antinodal direction and present data from MgB$_2$.

Our results  demonstrate that in a prototypical, conventional superconductor MgB$_2$, the coupling of the conduction electrons to phonon mode is not significantly affected by the SC transition. This is in contrast to behavior of the collective mode at antinode in cuprates, which disappears upon transition to normal state. We also discovered a signature of second collective mode in MgB$_2$ with energy $\Omega\sim$ 3.5 meV  that exists only below T$_c$. All characteristic of this mode are consistent with Leggett mode which arises due to the relative oscillation of the phases of two superconducting condensates present in this material.

We thank Andrey Chubukov, Mohit Randeria, Rafael Fernandes, Ilya Eremin and Mike Norman for very useful discussions. Research was supported by the US Department of Energy, Office of Basic Energy Sciences, Division of Materials Sciences and Engineering. Ames Laboratory is operated for the US Department of Energy by the Iowa State University under Contract No. DE-AC02-07CH11358. Work at Brookhaven (Bi2212 sample growth and characterization) is supported by the US DOE under Contract No. DE-AC02-98CH10886. JSW and ZJX are supported by the Center for Emergent Superconductivity, an Energy Frontier Research Center funded by the US DOE, Office of Science.

$^{*}$Corresponding author
kaminski@ameslab.gov

\newpage

{\bf \noindent \fontsize{18}{24} \selectfont Supplemental materials for Strong interaction between electrons and collective excitations in multiband superconductor MgB$_2$}

\section{Simulation procedure}

ARPES is an ideal tool with which to examine electron-boson coupling, as both the real, $\Sigma'(\omega)$ and imaginary, $\Sigma''(\omega)$ parts of the self-energy can be straightforwardly extracted from momentum distribution curves (MDCs). $\Sigma'(\omega)$ is the shift away from the bare dispersion, as extrapolated from high energy data, while the scattering rate $\Sigma''(\omega)$ is proportional to the width of the MDC.  For interactions with a collective mode at a single frequency, $\Omega$,  $\Sigma'(\omega)$ is peaked close to $\Omega$, while $\Sigma''(\omega)$ exhibits a step-like increase with the mid point close to $\Omega$, as quasiparticles can now decay into the mode.  The electron-boson coupling strength, $\lambda$ for each band depends on the renormalization of the effective mass, $\lambda = m^*/m -1$.  For a more complicated phonon spectrum, both parts of the self-energy can simply be calculated from the Eliashberg coupling function, $\alpha^2 F(\omega)$, which can be calculated numerically, allowing for a precise comparision.  The temperature dependence of $\Sigma(\omega)$ should be weak for $T \ll \Omega$.  However, as superconductivity gaps out the quasiparticles, the peak/step is expected to shift to $\Omega + \Delta$, at least within a simple BCS picture.  Superconductivity can similarly affect the energy of the dispersion anomalies, as phonons with energies above (below) the gap are expected to harden (soften) when the material is cooled through the transition.  Strong electron-electron interactions can also strongly renormalize the bandstructure, however, these effects should be small in MgB$_2$.

Below we illustrate the impact of features in the self energy on ARPES spectra by performing  simulations of the ARPES intensity for several scenarios of self energy. We start with an assumed form of the imaginary part of the self energy. We then calculate the real part of the self energy using a Kramer-Kronig transformation, using a reasonable value of the cut-off frequency $\omega_c$ of 1eV. The spectral function is then calculated using the following, well established formula:

\begin{equation*}
A({\bf k},\omega) =
{{|u_k|^2 \Sigma"} \over
{\left({\omega-\Sigma'-\sqrt{\epsilon_{\bf k}^2+\Delta^2}}\right)^2}+\Sigma"^2}
+ {{|v_k|^2 \Sigma" \over {\left({\omega-\Sigma'+\sqrt{\epsilon_{\bf k}^2+\Delta^2}}\right)^2}+\Sigma"^2}}
\end{equation*} 

where k is momentum, $\omega$ is binding energy, $\Sigma'$ and $\Sigma"$ are the real and imaginary parts of the self energy respectively, $\Delta$ is the magnitude of the superconducting gap, $\epsilon_{\bf k}$=v$_F$k is assumed linear band dispersion with Fermi velocity of 4 eV$\AA^{-1}$. Coefficients u$_k$ and v$_k$ are standard BCS coherence factors:

\begin{equation*}
|u_k|^2 ={1\over2}\left(1+{\epsilon_{\bf k}\over{\sqrt{\epsilon_{\bf k}^2+\Delta^2}}}\right)
\end{equation*} 

\begin{equation*}
|v_k|^2 ={1\over2}\left(1-{\epsilon_{\bf k}\over{\sqrt{\epsilon_{\bf k}^2+\Delta^2}}}\right)
\end{equation*} 

The ARPES intensify is then calculated by multiplying the spectral function by the Fermi function: $I({\bf k},\omega)=A({\bf k},\omega) f(\omega)$. 

\section{Results}

\subsection{Normal state}

We begin by plotting in Fig. S1 the calculated ARPES intensity and EDCs for the normal state assuming the form of self energy based on the data in Fig. 1 of the main text. The resulting dispersion is very similar to what is observed in the actual data, with a well pronounced kink due the changes in the real part of the self energy. Such changes in the real system are usually caused by the onset of scattering.

\subsection{Superconducting state}

In Fig. S2 we plot the ARPES intensity and dispersion calculated with a 6 meV superconducting gap and self energy of a Fermi liquid ($\Sigma$"=a+b$\omega^2$). The back bending of the band close to the E$_f$ is clearly observed, which is a characteristic signature of Bogolyubov quasiparticles. In Fig. S3 we plot the band dispersion and EDCs  in the presence of a 6 meV superconducting gap for a more realistic model of the self energy (similar to Fig. S1 but with a reduced offset in the imaginary part to better show the features in the proximity of E$_f$). The Bogolliubov-like dispersion can be still observed even though the peaks remain rather broad close to E$_F$.

\subsection{Dispersing vs non-dispersing quasiparticle peaks}

The appearance of a sharp quasiparticle peak in the data upon cooling below Tc (Fig. 3 of the manuscript) signifies a suppression of the scattering rate below a certain energy. We attempt to simulate this in Fig. S4, where we include a 6 meV superconducting gap and a simple suppression of the scattering rate below 10 meV. This results in a rather abrupt sharpening of the quasiparticle peak, but the peaks still display Bogoliubov dispersion, unlike what is observed in the experimental data where the sharp peak at the edge of the gap does not disperse. This lack of the dispersion requires a "resonant" feature in the scattering i. e. a peak in the imaginary part, in addition to the suppression of scattering. We illustrate this in Fig. S5, which is based on a similar self energy to Fig. S4, but with an additional peak in the imaginary part. This peak causes a more rapid change in the real part of the self energy, which in turn confines the sharp peak to a small energy range and significantly reduces its dispersion. This is of course a grossly exaggerated picture just to demonstrate the effect of peaks in the imaginary part of the self energy on the dispersion of the quasiparticles. A more realistic model is shown in Fig. S6, where we use the normal state self energy (from Fig. S1) that is suppressed below 10 meV and add a small gaussian peak at that energy. The resulting APRES intensity and EDC's very closely resemble the experimental data of Fig. 3 in main text. The band dispersion reaches an energy of -10 meV, above which only a sharp non-dispersing feature is observed. EDCs show a characteristic sharpening of the line shape with the sharp quasiparticle peak sitting at the superconducting gap energy. This peak is almost non-dispersing, just like in the data of Fig. 3. The line shape close to k$_F$ displays a characteristic hump-dip-peak structure, with a dip located close to 10 meV where the imaginary part of self energy changes rapidly.

\begin{figure*}[htbp]
\centering
\includegraphics[width=6in]{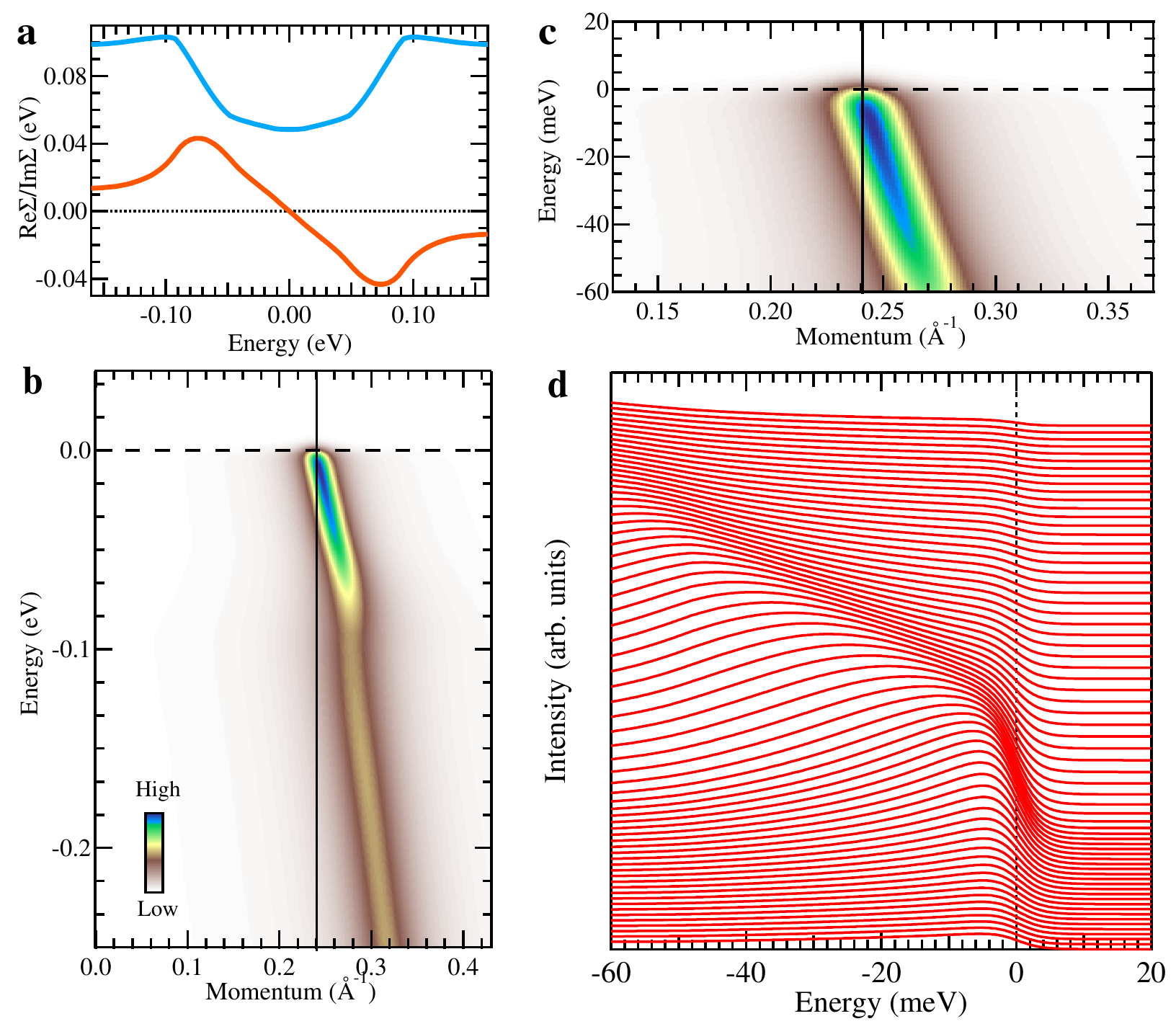}
\caption{(Color online) Simulated spectral function in the normal state in the presence of a collective mode (e. g. phonon) in normal state. a) The imaginary part of the self energy is similar to one extracted from the data in Fig. 1d of the main text. The real part of the self energy was calculated using a Kramers-Kronig transformation. b) The spectral function for the self energy in a) is multiplied by the Fermi function. c) same as (b) in close proximity to E$_f$. d) EDCs in the proximity of the Fermi wave vector. }
\end{figure*}

\begin{figure*}[htbp]
\centering
\includegraphics[width=6in]{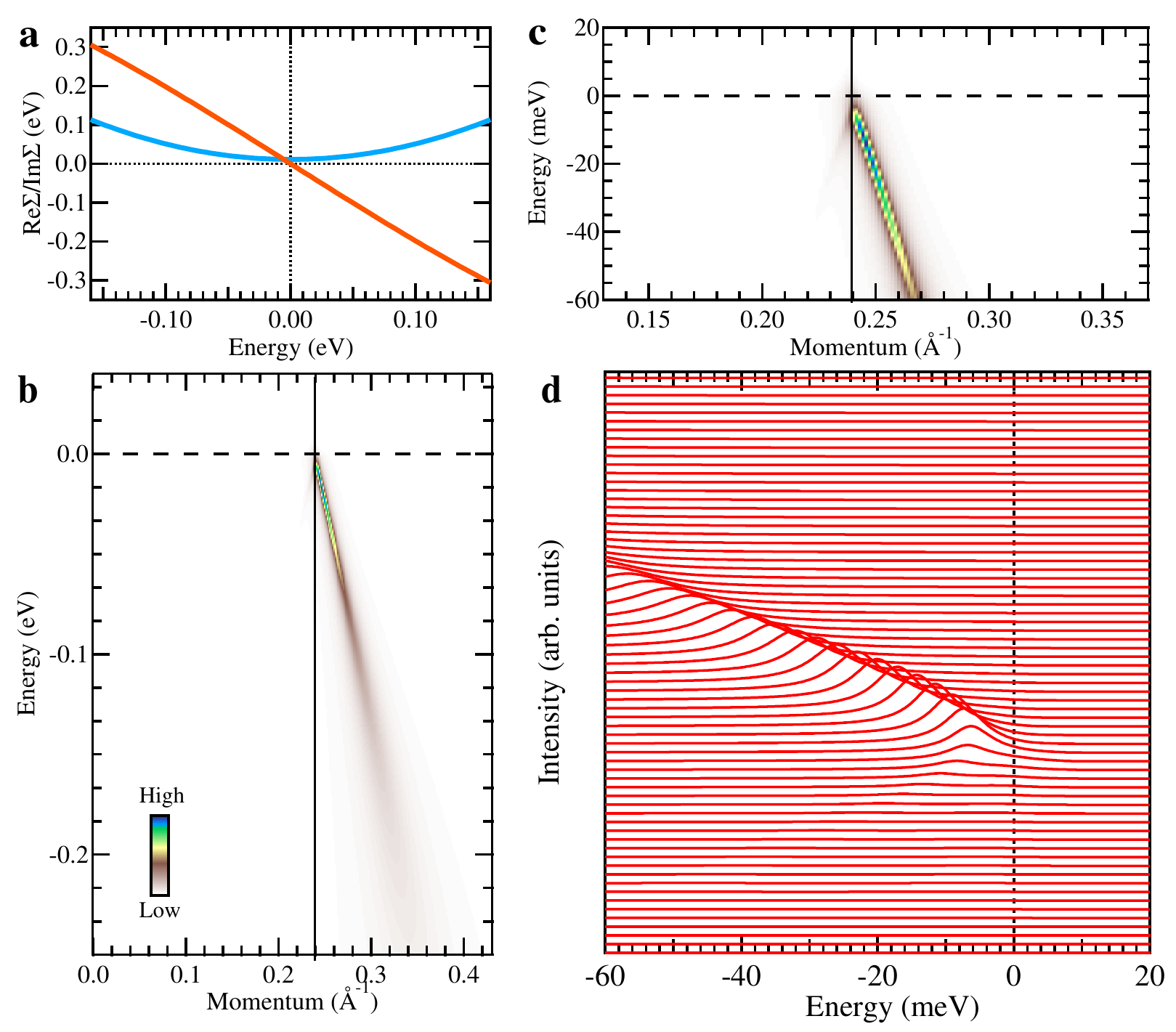}
\caption{(Color online) Simulated spectral function in the superconducting state with a gap of 6 meV and the self energy of a Fermi Liquid model. a)  The imaginary part of the self energy is proportional to $\omega^2$ with a small offset. The real part of the self energy was calculated using a Kramers-Kronig transformation. b) Spectral function for the self energy in a) is multiplied by the Fermi function.  c) same as (b) but plotted near E$_f$. d) EDCs in the proximity of the Fermi wave vector. Note that the peaks start to develop negative dispersion beyond k$_F$, a key signature of Bogolyubov quasiparticle peaks. They lose intensity away from k$_F$ due to coherence factors.}
\end{figure*}

\begin{figure*}[htbp]
\centering
\includegraphics[width=6in]{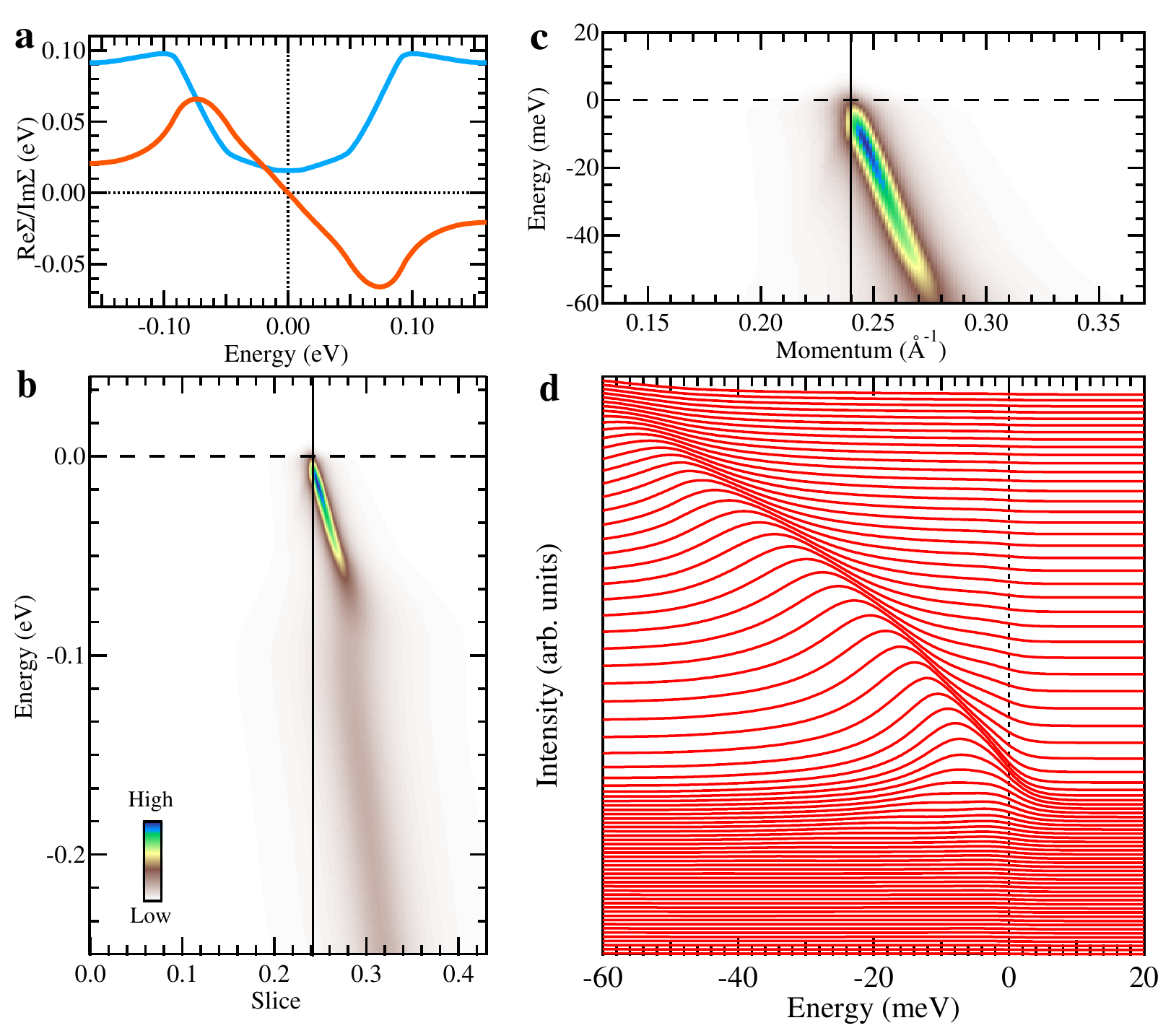}
\caption{(Color online) Simulated spectral function in the superconducting state with a gap of 6 meV and self energy similar to one in Fig 1, but with a smaller offset. a)  The imaginary part of the self energy is similar to Fig. 1a.  The real part of the self energy was calculated using a Kramers-Kronig transformation. b) Spectral function for the self energy in a) is multiplied by the Fermi function.  c) same as (b) but plotted near E$_f$. d) EDCs in the proximity of the Fermi wave vector. Bogolyubov dispersion is still visible even though the value of Im$\Sigma$ close to E$_F$ is large.}
\end{figure*}

\begin{figure*}[htbp]
\centering
\includegraphics[width=6in]{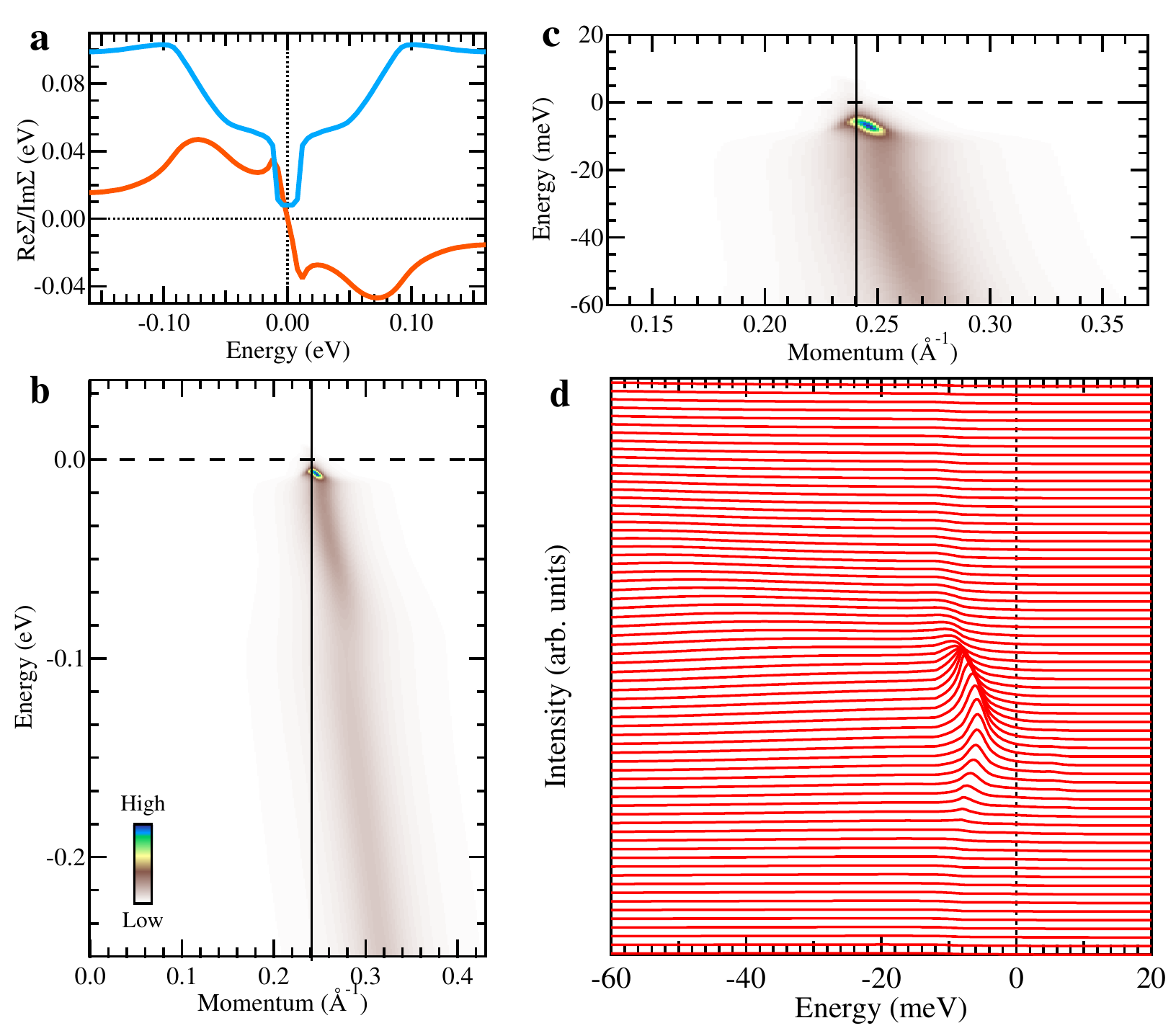}
\caption{(Color online) Simulated spectral function in the superconducting state with a gap of 6 meV and self energy that is strongly suppressed  below 10 meV. a)  Imaginary part of the self energy.  The real part of the self energy was calculated using a Kramers-Kronig transformation. b) Spectral function for the self energy in a) is multiplied by the Fermi function.  c) same as (b) but plotted near E$_f$. d) EDCs in the proximity of the Fermi wave vector. Bogolyubov dispersion is clearly observed above 6meV.}
\end{figure*}

\begin{figure*}[htbp]
\centering
\includegraphics[width=6in]{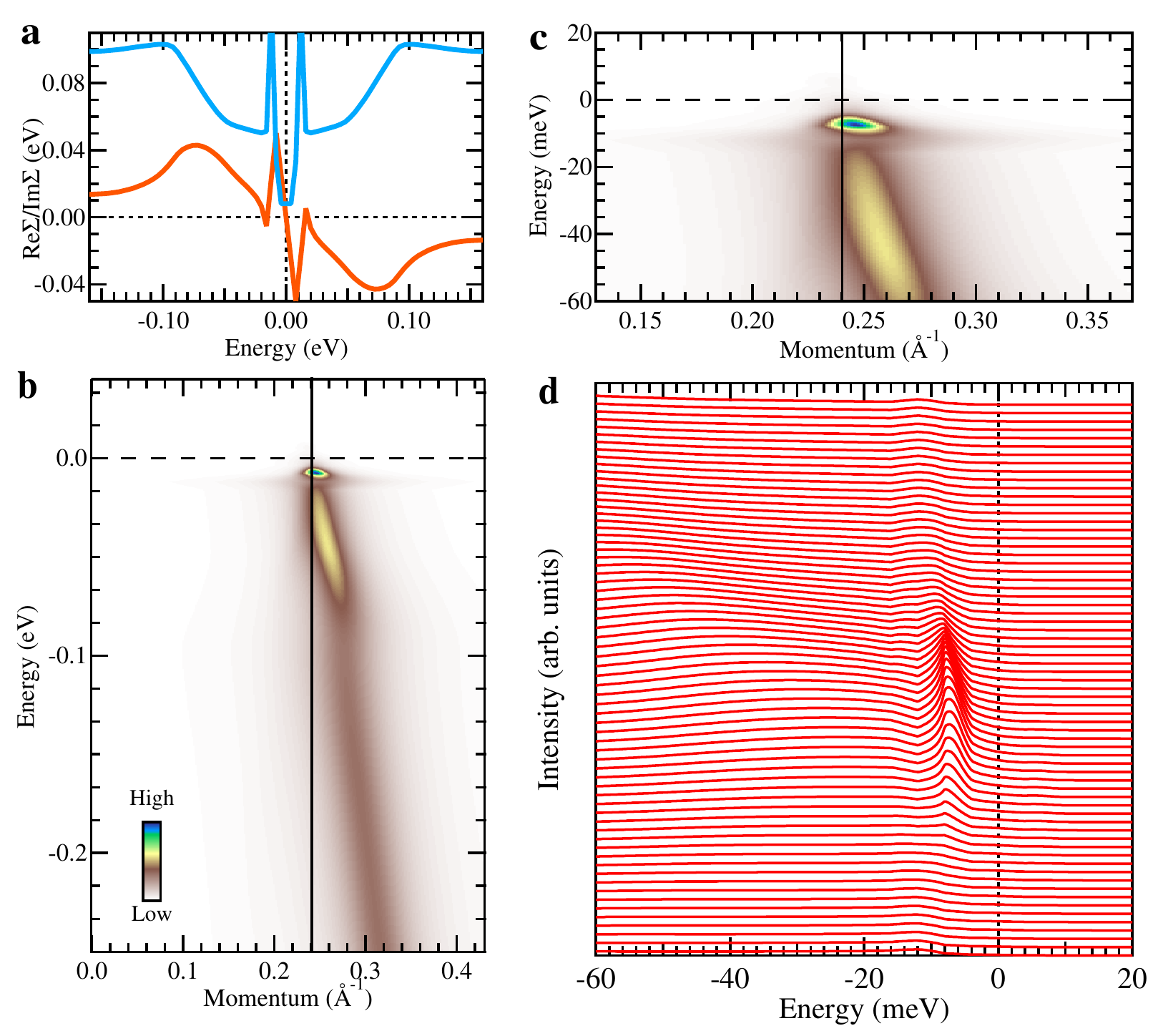}
\caption{(Color online) Simulated spectral function in the superconducting state with a gap of 6 meV and exaggerated self energy that is strongly suppressed  below 10 meV with an additional large gaussian peak at 10 meV. a)  Imaginary part of the self energy.  The real part of the self energy was calculated using a Kramers-Kronig transformation. b) Spectral function for the self energy in a) is multiplied by the Fermi function.  c) same as (b) but plotted near E$_f$. d) EDCs in the proximity of the Fermi wave vector. Bogoliubov dispersion is not observed, as the rapidly changing real part of the self energy confines the sharp quasiparticle peak.}
\end{figure*}

\begin{figure*}[htbp]
\centering
\includegraphics[width=6in]{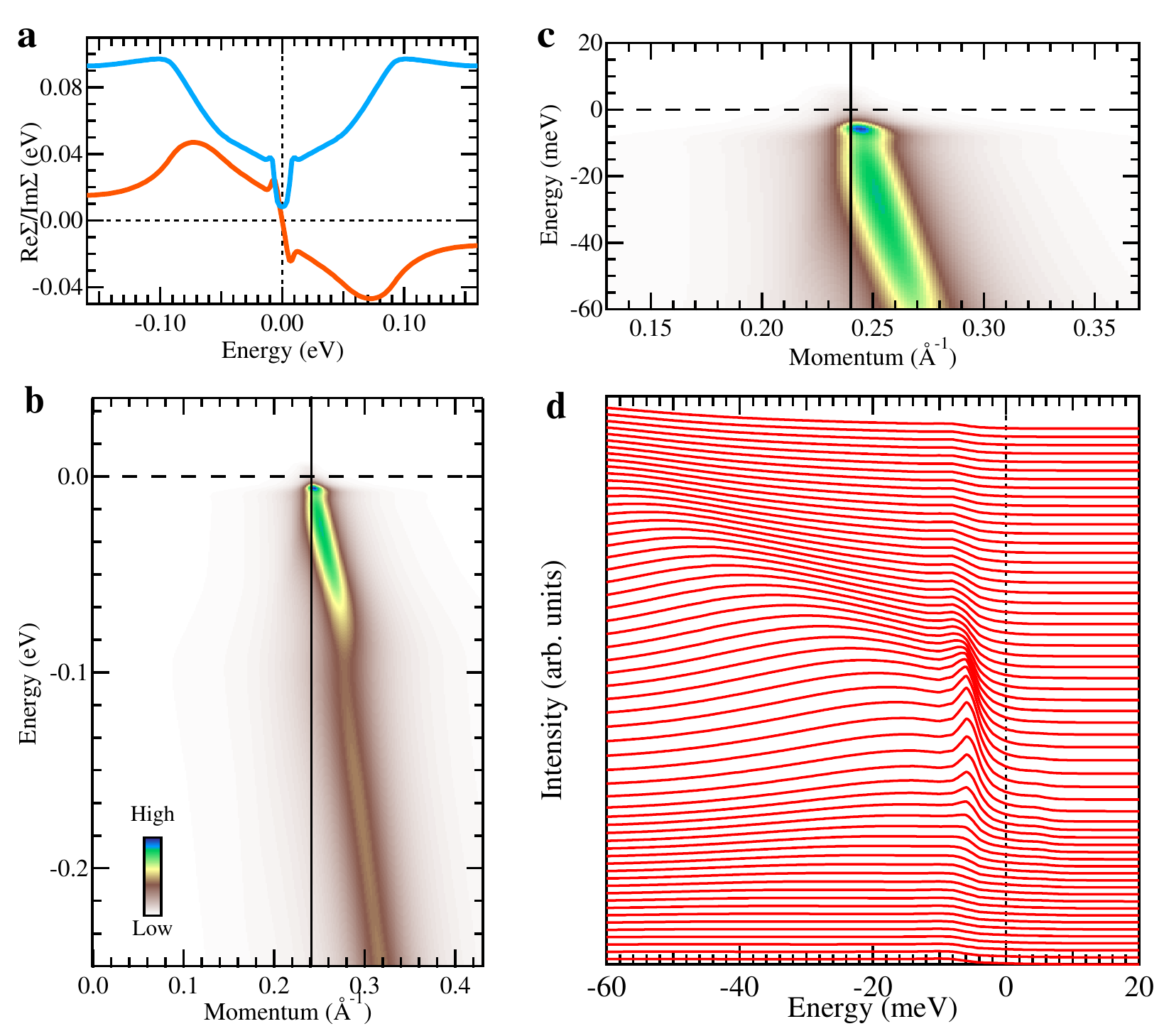}
\caption{(Color online) Simulated spectral function in the superconducting state with a gap of 6 meV and  a realistic self energy based on data from Fig. 1d of the main text, but suppressed  below 10 meV and with a small gaussian peak at that energy. a)  Imaginary part of the self energy.  The real part of the self energy was calculated using a Kramers-Kronig transformation. b) Spectral function for the self energy in a) multiplied by the Fermi function.  c) same as (b) but plotted near E$_f$. d) EDCs in the proximity of the Fermi wave vector. This simulation looks very similar to data in Fig. 3. The sharp peak present at the superconducting gap value is almost non-dispersing. The lineshape has a characteristic hump-peak-dip structure.}
\end{figure*}

\end{document}